# Analysis Of $SnS_2$ Buffer Layer And SnS Back Surface Layer Based CZTS Solar Cells Using SCAPS

Atul Kumar[1,a)] and Ajay D Thakur[1,2,b)]

[1]*Center for Energy and Environment, Indian Institute of Technology Patna, Patna 800013, India.*
[1]*Center for Energy and Environment, Indian Institute of Technology Patna, Patna 800013, India.* [2]*School of Basic Sciences, Indian Institute of Technology Patna, Patna 800013, India.*

[a)]atul.pph14@iitp.ac.in
[b)]ajay.thakur@iitp.ac.in

**Abstract.** A Copper-Zinc-Tin-Sulphide (CZTS) based solar cell with a modified cell configuration of Mo/SnS/CZTS/$SnS_2$/ZnO is simulated using SCAPS. An $SnS_2$ buffer layer is used in simulation instead of the standard CdS layer. An additional back surface passivation layer of SnS is added in the modified cell configuration. An improvement in the solar cell efficiency compared to the standard CdS buffer based solar cell configuration Mo/CZTS/CdS/ZnO is found. The observations suggest the possibility of using $SnS_2$ as a potential replacement of CdS. In addition, the use of SnS as a back surface passivation layer leads to improved solar cell performance.

## 1 INTRODUCTION

The per capita energy consumption of India is currently over 20 million BTU per annum. Owing to its growth trajectory, it is expected to increase further in near future. Due to limited reserves of coal and oil and an over increasing ecological concern related with hydroelectricity and nuclear power, we will need alternative sources of energy to meet this ever increasing demand. Solar energy or photovoltaics (PV) is omnipresent, free of cost, and eco-friendly thereby presents a viable alternative. Silicon based solar cells require intensive infrastructure for massive production and hence alternate PV technologies are being explored the world over. Among various other PV technologies, thin film solar cells based on CZTS can be a potential candidate owing to its non toxic earth abundant constituent and high detailed balance limit of 32.4% for its efficiency [1]. Highest laboratory efficiency reported till date is 12.6% [1] for CZTS/CdS hetrojunction whereas efficiency of CIGS/CdS has reached 21.6 % [2]. Although CIGS has high efficiency but scare Indium diminishes its prospect for large scale PV. The lower efficiency of In replaced CZTS has forced researchers to reexamine it. This highly promised material currently have not achieved efficiency comparable to CIGS due to possible challenge of defect-less absorber layer fabrication, non optimized buffer layer matching and non optimized cell configuration. The current configuration of CZTS solar cell is similar to well established CIGS or CIS technology. The use of window layer is not limited to isolate device from outer contamination [3] but also to provide low resistive path to contacts which can be optimized further. The p-type CZTS absorber is fabricated from wide range of techniques from vacuum deposition methods to direct liquid coating with dedicated effort to optimize it [4].The cell with different buffer layers can be analyzed by simulation software, like SCAPS, wxAMPS, PC1D etc [5] where various cell structure/design with different buffer layers can be simulated and optimized. Some suitable nontoxic buffer layer like ZnO, ZnS, ZnSe, ZnMgO, $InZnSe_x$, $SnO_2$, $In_2S_3$

[6], TiO$_2$ [7] are simulated for a possible alternative of characteristically hazardous CdS buffer in both CZTS and CIGS. In this regard we propose n-type SnS$_2$ layer as a possible buffer layer for CZTS thin film solar cell. A back surface passivation layer of SnS is added to the absorber layer to modify the overall cell configuration. This particular p-SnS/p-CZTS/n-SnS$_2$/ZnO is simulated with SCAPS-1D software to study and optimize the various parameter of device like bandgap, thickness of layers, recombination and overall performance. A comparative study of Solar cell with modified configuration of Mo/SnS/CZTS/SnS$_2$/ZnO and standard Mo/CZTS/CdS/ZnO is made which are further discussed in next section. A simulation, studying the effect of various parameters on photovoltaic performance is an easy step to predict optimizing condition without actually fabricating and characterizing a solar cell. Highly sought after SCAPS-1D simulation software is used in this study solves Poisson and Continuity equation at the boundary for simulation of devices [8]. SCAPS can simulate cell structures of the CuInSe$_2$ and the CdTe family, crystalline solar cells (Si and GaAs family) and amorphous cells (a-Si and micromorphous Si)[8], Pervoskite [9] and CZTS[10].

## 2 CELL STRUCTURE AND SIMULATION DETAILS

ZnS, ZnMgO, In$_2$S$_3$, TiO$_2$ are the some of the material tried for replacement of CdS buffer layer owing to their availability non toxicity and feasibility. SnS$_2$ is proposed as buffer layer for CZTS for the first time as per our knowledge, since it has similar Bandgap value as that of CdS, and is n-type material [11]. It can be easily synthesized by various wet chemical [11] and solvothermal techniques. The cell simulated with SnS back surface passivation layer and SnS$_2$ buffer layer is called modified cell and compared with standard cell. The structure of modified cell is Mo/SnS/CZTS/SnS$_2$/ZnO and that of Standard cell is Mo/CZTS/CdS/ZnO. Schematic of both the cells are shown in figure (1) and are simulated using SCAPS-1D version 3.3.02. The material parameters used are listed in the Table 1. SnS BSL, CZTS absorber, SnS$_2$ buffer layer, ZnO as window layer and front contact, Mo taken for Back contact. An elaborate table with various Cell configurations and their performance is presented in the end of this paper. The device is illuminated from window layer side with AM1.5 spectrum from SCAPS illumination value. The effect of various parameters on FillFactor, efficiency is analyzed by SCAPS using its batch option. Normal operating conditions are temperature 300 Kelvin, series resistance of 4.25Ω-cm$^2$, shunt resistant of 400 Ω-cm$^2$. SnS, CZTS, SnS$_2$ and window layer thickness and other parameter are given in the Table 1. The Mo back contact of work function 5eV is used in the simulation.

**TABLE 1.** Material parameters used in simulation.

| Parameters | CZTS | SnS$_2$ | SnS | ZnO |
|---|---|---|---|---|
| Thickness(nm) | 2680 | 100 | 20 | 100 |
| Bandgap(eV) | 1.5 | 2.24 | 1.25 | 3.3 |
| Electron affinity(eV) | 4.5 | 4.24 | 4.2 | 4.6 |
| Electron thermal velocity(cm/s) | $10^7$ | $10^7$ | $10^7$ | $10^7$ |
| Electron mobility(cm$^2$/Vs) | $10^2$ | 50 | 25 | $10^2$ |
| Donor density N$_D$ (cm$^{-3}$) | 10 | $1*10^{17}$ | 10 | $10^{18}$ |
| Acceptor density N$_A$ (cm$^{-3}$) | $1*10^{17}$ | 10 | $3*10^{19}$ | 10 |
| Absorption coefficient(cm$^{-1}$) | $2*10^4$ | $5*10^3$ | $10^4$ | SCAPS value |

## 3 RESULT AND DISCUSSION

### 3.1. Buffer layer SnS$_2$

A p-n Hetrojuntion is formed by employing n-type SnS$_2$ as a buffer layer. SnS$_2$ may absorb light leading to less number of photon available to be absorbed in absorber thus less e-h pair. Keeping other parameters constant Absorber layer (t$_{abs}$=2.68 um), an analysis is done to study the effect of buffer layer thickness on efficiency which shows optimum SnS$_2$ buffer layer thickness around t$_{buffer}$ = 100nm. Cell efficiency decrease for higher buffer layer thicknes as shown in the figure

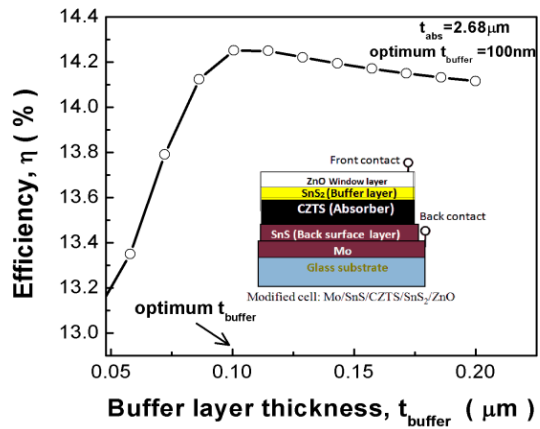

**FIGURE 1.** Effect of Buffer layer thickness on efficiency of modified cell. Inset shows a schematic diagram of the modified cell with SnS back surface layer and $SnS_2$ buffer layer.

## 3.2. Absorber layer

The CZTS absorber layer has high absorption coefficient for visible light, a suitable direct bandgap of 1.5eV, p type conductivity. The p type nature is due to the stoichiometric defect of Cu valency, and Cu-on-Zn antisites $Cu_{Zn}$ [12]. The Grading of Cu, Zn and Sn in absorber layer can be tuned to obtained high performance in a solar cell. It is reported that Cu poor zinc rich CZTS are good for PV applications. Absorber layer properties are most critical for overall cell performance. Bandgap of CZTSSe can be varied with S/Se ratio from CZTS 1.5eV to CZTSe 1eV. The S/Se ratio leads to variation in hole density, bandgap and other properties of CZTS. The CZTS absorber can be optimized by varying Cu/Zn+Sn, S/Se ratio to set material properties and band gap. CZTS bandgap can be varied so as to match with solar spectrum for improved performance of cell. A width of absorber layer is required for light absorption. Absorption coefficient determines minimum thickness needed for complete absorption. CZTS has absorption coefficient in order of $10^4$ cm$^{-1}$. Efficiency grows from 0 to 12% for 0-2 μm absorber thickness, in next 1um it only grows by 2 % as is shown in simulated curve. To get optimum performance output and cost effectiveness a minimum thickness is required. The absorber thickness $t_{abs}$ is taken as 2.68 micrometer in this simulation. This is done so as to compare our simulation results with the ones without SnS back surface layer. The simulation curve shows the optimum thickness for CZTS in 2.5-3 micrometer range when keeping $t_{buffer}$ at optimum thickness of 100 nm. The fill-factor (FF) initially increases then becomes a constant with thickness. FF depends upon the shunt and series resistance values of the cell.

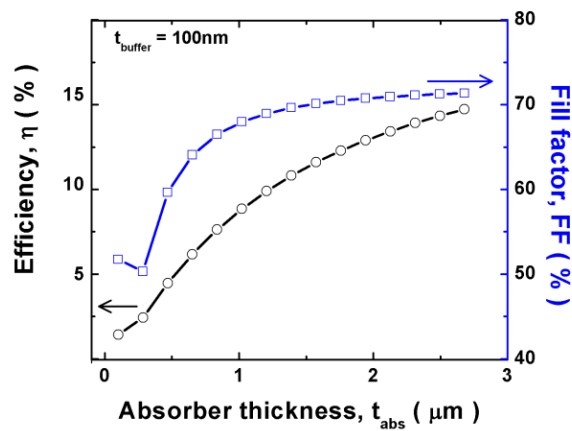

**FIGURE 2.** Absorber layer (CZTS) thickness dependent variation of efficiency and Fillfactor.

## 3.3. Recombination current

Lower recombination current is observed in the Modified cell with SnS back surface layer. The difference of recombination current is of the order of 10 in the two cells. The variation of recombination with thickness of CZTS absorbers are simulated and are compared for both standard cell and modified cell as shown in figure. Adding a SnS layer has reduced the recombination current in modified cell. A conductive path for hole is provided by the SnS layer may be the cause of reduction in recombination CZTS bulk. A back surface layer of lower work function provide hole conducting path at interface of CZTS and back surface. Adding a SnS layer at back surface lower the hole barrier by band banding at interface in modified cell thus reducing the effective hole density in absorber and the chance of electron–hole recombination in bulk. A material of low work function and suitable band gap will act as a hole conducting and electron blocking back surface layer.

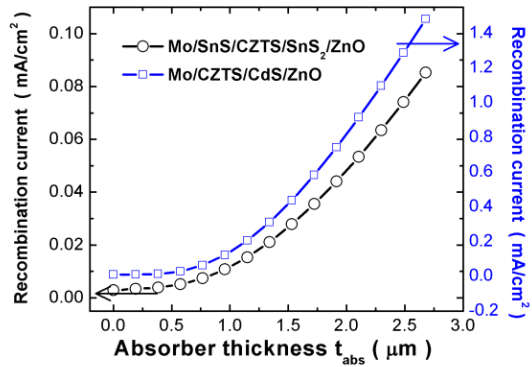

**FIGURE 3.** Comparison of Recombination current in modified and standard cell.

## 3.4. Optimized Performance of Mo/SnS/CZTS/SnS$_2$ solar cell

The final simulated IV curve is shown. The modified cell has showed the better and higher efficiency than the standard cell. The p type SnS layer can prove beneficial for cell performance. The overall performance is listed in table 3 and compared with Reference 11 and 15.

**TABLE 2.** Comparison of Modified Cell with References.

| Cell configuration | Efficiency (%) | FillFactor(%) | Voc(V) | Jsc (mA/cm$^2$) | Reference |
|---|---|---|---|---|---|
| Mo/CZTS/CdS/ZnO | 6.44 | 62.89 | .589 | 17.6 | 10 |
| BM/CZTS/CdS/ZnO | 13.41 | 69.35 | 1.002 | 19.3 | 10 |
| Mo/CZTS/CdS/ZnO | 8.97 | 63.41 | .6407 | 22.0721 | 13 |
| Mo/SnS/CZTS/SnS$_2$/ZnO | 14.249 | 71.33 | .992 | 20.1315 | This work |

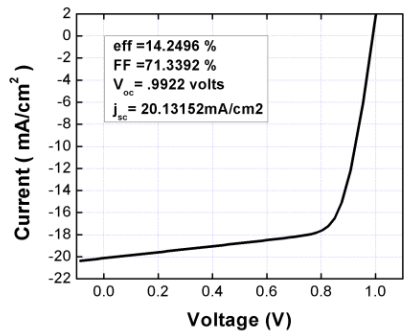

**FIGURE 4.** IV curve of Modified Mo/SnS/CZTS/SnS$_2$/ZnO solar cell.

# 4 CONCLUSIONS

The modified CZTS cell with $SnS_2$ buffer layer and SnS Back surface layer have been proposed and various Curves have been analyzed using SCAPS. Analysis showed modified design has higher simulated performance than the standard cell. Adding SnS Back surface layer has reduced the recombination in absorber layer. An actual experimental implementation for verification can be done to achieve this better performance. The simulation has explored $SnS_2$ as potential replacement of CdS along with improvement in solar cell efficiency. The versatile tin sulfide family SnS is p type material and $SnS_2$ is n type material. Together they can be employed in solar cell with improved performance.

# ACKNOWLEDGMENT

The solar cell capacitance simulator SCAPS-1D used here for simulation is provided by Dr. Marc Burgelman from the University of Gent.